\documentclass[pra,floatfix,footinbib,reprint,superscriptaddress]{revtex4-1}

\usepackage{iopams,setstack,amsopn} 
\usepackage{amsmath, amssymb, amsthm, amsfonts}
\usepackage{graphicx}
\usepackage{dcolumn}
\usepackage{bm}
\usepackage{times}
\usepackage{subcaption}
\usepackage{dsfont}
\usepackage{tablefootnote}
\usepackage{color}
\usepackage{verbatim}
\usepackage[pdftex, colorlinks, allcolors=blue]{hyperref}
\usepackage[]{ragged2e}
\usepackage{soul}
\newcommand{\keta}[1]{| #1_{a} \rangle}
\newcommand{\ketb}[1]{| #1_{b} \rangle}
\newcommand{\ketmu}[1]{| #1_{\mu} \rangle}
\newcommand{\bramu}[1]{\langle #1_{\mu} |}

\newcommand{\braa}[1]{\langle #1_{a} |}

\newcommand{\bracketba}[2]{\langle #1_{b} | #2_{a} \rangle}

\begin{document}

\title{
Experimental verification of the treatment of time-dependent flux in circuit quantization
}

\author{Jacob Bryon}
\affiliation{Department of Electrical Engineering, Princeton University, Princeton, 
New Jersey, 08540, USA}
\author{D.~K.~Weiss}
\affiliation{Department of Physics and Astronomy, Northwestern University, Evanston, Illinois 60208, USA}
\author{Xinyuan You}
\email[Present address: Superconducting Quantum Materials and Systems Center, Fermi National Accelerator Laboratory (FNAL), Batavia, IL 60510, USA]{}
\affiliation{Graduate Program in Applied Physics, Northwestern University, Evanston, Illinois 60208, USA}
\author{Sara Sussman}
\affiliation{Department of Physics, Princeton University, Princeton, New Jersey, 08540, USA}
\author{Xanthe Croot}
\affiliation{Department of Physics, Princeton University, Princeton, New Jersey, 08540, USA}
\author{Ziwen Huang}
\email[Present address: Superconducting Quantum Materials and Systems Center, Fermi National Accelerator Laboratory (FNAL), Batavia, IL 60510, USA]{}
\affiliation{Department of Physics and Astronomy, Northwestern University, Evanston, Illinois 60208, USA}
\author{Jens Koch}
\affiliation{Department of Physics and Astronomy, Northwestern University, Evanston, Illinois 60208, USA}
\affiliation{Center for Applied Physics and Superconducting Technologies, Northwestern University, Evanston, Illinois 60208, USA}
\author{Andrew Houck}
\affiliation{Department of Electrical Engineering, Princeton University, Princeton, 
New Jersey, 08540, USA}

\date{\today}

\begin{abstract}

Recent theoretical work has highlighted that quantizing a superconducting circuit in the presence of time-dependent flux $\Phi(t)$ generally produces Hamiltonian terms proportional to $d\Phi/dt$ unless a special allocation of the flux across inductive terms is chosen. Here, we present an experiment probing the effects of a fast flux ramp applied to a heavy-fluxonium circuit. The experiment confirms that na\"ive omission of the $d\Phi/dt$ term leads to theoretical predictions inconsistent with experimental data. Experimental data are fully consistent with recent theory that includes the derivative term or equivalently uses ``irrotational variables'' that uniquely allocate the flux to properly eliminate the $d\Phi/dt$ term.

\end{abstract}

\maketitle

\section{Introduction}


The use of time-dependent flux is ubiquitous in the context of circuit quantum electrodynamics (cQED). Flux modulation is utilized for, e.g., performing fast two-qubit gates with transmons \cite{Beaudoin2012,Strand2013, McKay2016, Mundada2019}, demonstrating universal stabilization of arbitrary single-qubit states between a qubit and a resonator \cite{Lu2017, Huang2018}, sweet-spot engineering of transmons using two-tone flux modulation \cite{Valery2022}, enhancing fluxonium coherence via Floquet engineering \cite{Mundada2020, Huang2020}, and performing fast gates on fluxonium qubits \cite{Zhang2021, Chen2021, Moskalenko2021, Moskalenko2022, Weiss2022}. Given the importance of flux modulation for the control of superconducting circuits, it is critical to accurately model time-dependent flux in circuit Hamiltonians.

Recent theory work has investigated the treatment of time-dependent external flux in superconducting circuits both for lumped-element models \cite{You2019}, and for distributed geometries \cite{Riwar2021, Rajmohan2022}. Considering the lumped-element model of the fluxonium circuit as an example, external flux is usually allocated to either the inductor term in the Hamiltonian \cite{Koch2009} or to the Josephson junction term \cite{Nguyen2019,Sete2017, Spilla2015}. Under static conditions, both  choices are related by a constant variable shift and naturally result in identical predictions. This ceases to be true in the context of time-dependent flux $\Phi(t)$. In the case of fluxonium, the junction allocation of time-dependent external flux yields an additional term in the Hamiltonian which is proportional to the time derivative of the external flux, $d\Phi/dt$ \cite{You2019}. By contrast, the inductor allocation of external flux produces the more familiar form of the Hamiltonian in which the $d\Phi/dt$ term is absent, making this special choice valid in both time-independent and time-dependent scenarios.

In this work, we experimentally test and compare predictions for the dynamics under time-dependent flux, based on the inductor allocation and the incomplete junction allocation in which the additional $d\Phi/dt$ term is incorrectly omitted. 
Beginning with a flux bias $\Phi_a$ at or near the half-flux sweet spot, we apply a fast flux pulse to ramp to a new flux point $\Phi_b$. We then read out the occupation probability in the ground and first-excited states, and compare the experimental data to theory based on the inductor allocation and the incomplete junction allocation. We find excellent agreement with theory based on the correct inductor allocation and poor agreement with the incomplete junction allocation. These findings provide experimental evidence for the theory of circuit quantization presented in Ref. \cite{You2019}, and highlight the importance of using irrotational variables when working with time-dependent flux.

\section{Theory}
\label{sect:Theory}

\begin{figure}
     \centering
     \includegraphics[width=0.43\columnwidth]{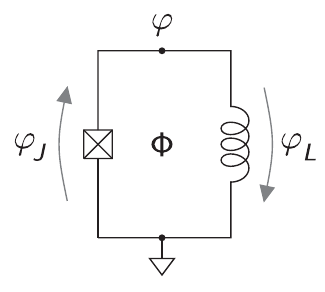}
     \caption{
        The fluxonium circuit is formed by a Josephson junction and an inductor with respective branch variables $\varphi_{J}, \varphi_{L}$. The bottom node is chosen as the reference ground while the top node is associated with the node variable $\varphi$. The junction and inductor form a loop which may be threaded by an external flux $\Phi$.
        }
     \label{fig:FluxoniumCircuit}
\end{figure}

In the literature, the fluxonium Hamiltonian is found in the following two forms~\cite{Koch2009,Nguyen2019, Sete2017, Spilla2015}:
\begin{align}
\label{eq:ham_ind}
H_1&=4E_{C}n^2-E_{J}\cos(\varphi)+\frac{1}{2}E_{L}(\varphi-\phi)^2,  \\ 
\label{eq:ham_jj}
H_2&=4E_{C}n^2-E_{J}\cos(\varphi+\phi)+\frac{1}{2}E_{L}\varphi^2.
\end{align}
Here, $E_{C}$, $E_{L}$, and $E_{J}$ denote the usual charging, inductive, and Josephson energies, and the phase and charge operators $\varphi$ and $n$ are canonically conjugate and obey $[\varphi,n]=i$. We have further defined the reduced external flux $\phi=2\pi\Phi/\Phi_{0}$ in terms of the applied magnetic flux $\Phi$ and superconducting flux quantum $\Phi_{0}=h/2e$. 
The key difference between Eqs.~\eqref{eq:ham_ind} and~\eqref{eq:ham_jj} is the allocation of the external flux $\phi$ to the inductive terms in the Hamiltonian: flux is allocated to the inductor term in $H_{1}$, but to the junction term in $H_2$.
When the external flux is static, both choices are equivalent \cite{Devoret1995, Vool2017}.
However, in the presence of time-dependent flux $\phi\to\phi(t)$, this equivalence no longer holds. 
Following Refs.~\cite{You2019,Riwar2021}, the proper handling of time-dependent flux must already occur at the level of the classical Lagrangian. 
In short, the fluxonium circuit involves two branch variables: the branch variable $\varphi_J$ of the junction and the branch variable $\varphi_L$ of the inductor, see Fig.~\ref{fig:FluxoniumCircuit}. 
Fluxoid quantization ~\cite{orlando1999superconducting} now imposes semi-holonomic constraints for both the generalized coordinates and the velocities \cite{Goldstein}:
\begin{equation} 
\label{eq:tdhc}
    \varphi_J + \varphi_L = \phi(t), \quad
    \dot{\varphi}_J + \dot{\varphi}_L = \dot{\phi}(t).
\end{equation}
From here, one proceeds by eliminating one of the variables and its time derivative. This leads to a Lagrangian describing a single degree of freedom which will generally involve a term proportional to $\dot{\phi}(t)$. One choice of variables  (also known as the irrotational variables in Ref.~\cite{You2019}) succeeds in the elimination of that term. For the fluxonium circuit this special choice leads to the Hamiltonian 
\begin{align}
\label{eq:H1t}
    H_{1}(t)=4E_{C}n^2-E_{J}\cos(\varphi)+\frac{1}{2}E_{L}[\varphi-\phi(t)]^2,
\end{align}
where the time-dependent external flux is fully allocated to the inductor term. By contrast, if we force the time-dependent external flux to be allocated with the junction, we obtain the Hamiltonian
\begin{align}
\label{eq:H2t}
H_{2}(t) &=4E_{C}n^2-2en\dot{\Phi} \\ \nonumber 
&\quad-E_{J}\cos[\varphi+\phi(t)]+\frac{1}{2}E_{L}(\varphi)^2.
\end{align}
This includes the aforementioned time-derivative term. Thus the na\"ive generalization of Eq.~\eqref{eq:ham_ind} for $\phi\rightarrow\phi(t)$ is correct [c.f. Eq.~\eqref{eq:H1t}], while simply taking $\phi\rightarrow\phi(t)$ in Eq.~\eqref{eq:ham_jj} is incorrect [c.f. Eq.~\eqref{eq:H2t}].
The central goal and main result of this paper are the experimental verification of this theory.

The effects of a fast flux pulse provide an ideal testbed for comparing predictions based on the inductor allocation vs.\ the incomplete junction allocation. (We will not discuss further the complete and correct form of junction allocation [Eq.~\eqref{eq:H2t}], which must give results identical to those obtained from the inductor allocation of the time-dependent flux.)
We devise a simple experiment based on dc flux-biasing the qubit near the half-flux sweet spot, then ramping quickly to a new value of flux, and finally performing readout. To see how the two allocation methods will yield different predictions, consider the following. At the initial and final flux bias points $\Phi_{a}$ and $\Phi_{b}$, respectively, we write the Hamiltonian in its eigenbasis
\begin{align}
H(\Phi_{\mu})=\sum_{j}\omega_{j, \mu}\ketmu{j}\bramu{j},\quad \mu=a,b.
\end{align}
At this point we need not specify junction or inductor allocation. While both allocations give the same results for eigenenergies and expectation values \cite{You2019}, the wave functions $\ketmu{j}$ do depend on the flux allocation, as we will discuss below.

\begin{figure}
     \centering
     \includegraphics[width=\columnwidth]{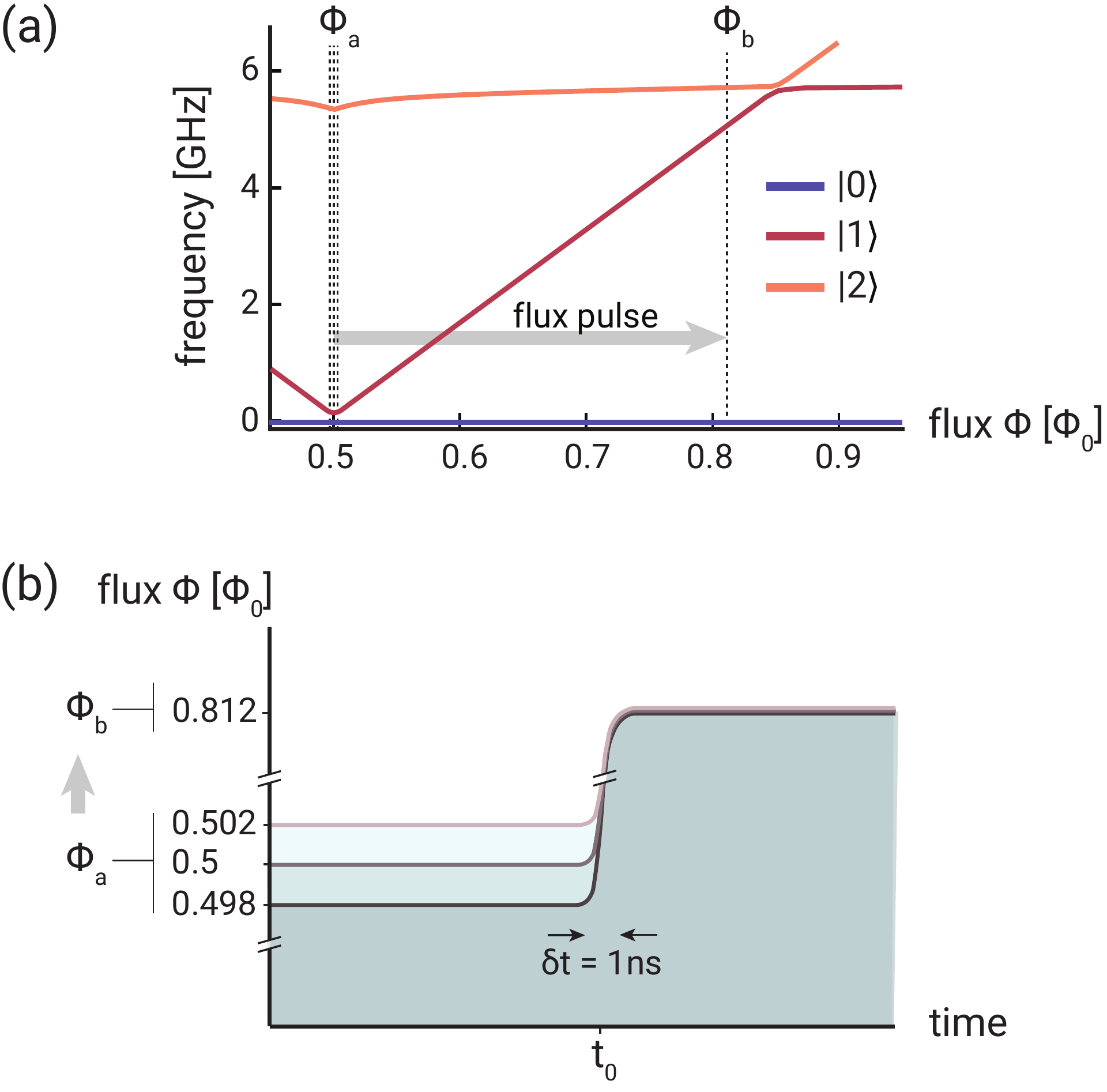}
     \caption{
         (a) Fluxonium energy spectrum as a function of external flux, showing the first three energy levels. Vertical lines indicate flux bias points used in the experiment. Starting flux biases are close to half-flux, $\Phi_a/\Phi_0\in [0.48, 0.502]$. The final flux bias reached after the pulse is $\Phi_b/\Phi_0 = 0.812$.
        (b) External flux as a function of time for the sudden approximation experiment. The dc flux of the external magnet can be precisely tuned to set the starting point near half-flux. Then, using the QICK qubit controller, we generate a step function with a $1\,\text{ns}$ rising edge to drive external flux via the on-chip flux bias line. We calibrate the drive power such that we always drive to $\Phi_b/\Phi_0 = 0.812$. By tuning the starting flux point we are able to change the initial state and thus the final state occupation probability.
        }
     \label{fig:FluxVsTime}
\end{figure}

\begin{figure*}[ht]
    \centering
    \includegraphics[width=2\columnwidth]{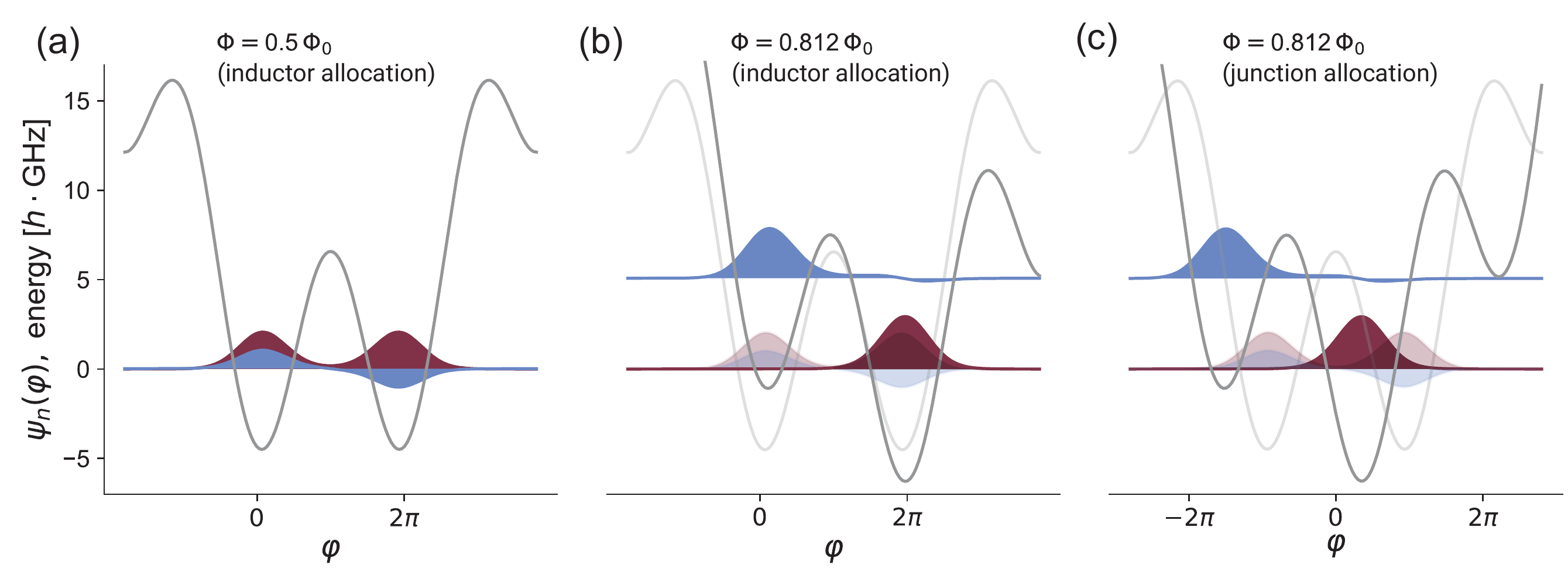}
    \caption{\label{fig:eigenstates} Fluxonium eigenstates as a function of flux. (a) The two lowest-energy fluxonium eigenstates at the half-flux sweet spot. (b-c) The two lowest-energy fluxonium eigenstates at $\Phi/\Phi_{0}=0.812$ using the (b) inductor and (c) junction allocation of the flux. The potential and eigenstates for half-flux are transparently overlaid. While the eigenenergies do not depend on the allocation of the external flux to different terms in the fluxonium Hamiltonian, the shifts of the potential minima and thus the relative wave function locations in $\varphi$ space do depend on it. Since final occupation probability depends on wave function overlaps, the incomplete junction allocation and inductor allocation produce different predictions. Circuit parameters used are $E_{J}/h=6.49\,\text{GHz}$, $E_{C}/h=0.755\,\text{GHz}$, $E_{L}/h=0.445\,\text{GHz}$. Wave function plots are generated using scqubits \cite{scqubits}.
    }
\end{figure*}

At time $t=t_{0}$ with the system in state $|\psi_{\mathrm{initial}}\rangle$, we imagine switching on a fast, non-adiabatic flux pulse which ramps from $\Phi_a$  to the final flux point $\Phi_{b}$ (Fig.~\ref{fig:FluxVsTime}). 
Within the sudden approximation \cite{sakurai}, the quantum state remains unchanged under a rapid  flux
pulse. The probability of occupying an $H(\Phi_{b})$-eigenstate $|m_{b}\rangle$ at times later than $t_{0}$ is
\begin{align}
\label{eq:prob_initial}
p_{m}=|\langle m_{b}|\psi_{\mathrm{initial}}\rangle|^2.
\end{align}
Crucially, the overlap in Eq.~\eqref{eq:prob_initial} is altered if the $d\Phi/dt$ term is erroneously omitted for the junction allocation of the external flux. 
This finding can be observed straightforwardly in Fig.~\ref{fig:eigenstates}. 
In Fig.~\ref{fig:eigenstates}(a) we show the qubit's lowest-lying wave functions at $\Phi=0.5\ \Phi_0$.  In Fig.~\ref{fig:eigenstates}(b)-(c) we compare the wave functions at $\Phi=0.812\ \Phi_0$ to those at $\Phi=0.5\ \Phi_0$ for inductor and junction allocations respectively. We observe that, e.g., the overlap of the ground state at $\Phi_{b}$ with the ground state at $\Phi_{a}$ in the case of inductor allocation is larger than for the incomplete junction allocation, due to the shift in minimum locations. Because the wave functions are centered around potential minima, the differing shifts in the minimum locations lead to quantitatively different results.

To see why the minimum shifts should differ based on the flux allocation, consider the following discussion of the change in minimum location based on a small flux shift. (The following analysis offers intuition but does not yield quantitatively accurate results for the experiment conducted here, as the flux shift here is not ``small.")
Consider the dc flux $\phi$ and the associated global minimum of the potential energy $\varphi=\bar{\varphi}$. (In the case of degenerate minima, select either one of them.) Suppose now we add a perturbation $\phi\rightarrow\phi+\delta\phi$. Solving for the perturbed minimum location $\bar{\varphi}+\delta\varphi$, we obtain in the case of inductor allocation
\begin{align}
\delta\varphi = \frac{E_{L}}{E_{L}+E_{J}\cos(\bar{\varphi})}\delta\phi\approx\frac{E_{L}}{E_{J}}\delta\phi,
\end{align}
and in the case of junction allocation
\begin{align}
\delta\varphi = \frac{E_{J}\cos(\bar{\varphi}-\phi)}{E_{L}+E_{J}\cos(\bar{\varphi}-\phi)}\delta\phi\approx \delta\phi.
\end{align}
We have neglected quadratic corrections, and the approximation holds in the parameter regime $E_{J}\gg E_{L}$ typical of fluxonium qubits. Thus, the shift is large for the junction allocation compared to the inductor allocation of the flux. 
(We emphasize that the complete junction allocation not omitting the $d\Phi/dt$ term yields the same results as the inductor allocation.)  

Choosing initial flux biases near half-flux is advantageous for two  reasons.  First, the nature of the eigenstates $\keta{j}$ changes dramatically in the vicinity of the half-flux point. For $\Phi/\Phi_0=0.5$ the qubit eigenstates are delocalized across the two degenerate minima, see Fig.~\ref{fig:eigenstates}. Moving away from half-flux, the eigenstates quickly localize in a single minimum. Thus, small shifts in the starting flux bias $\Phi_{a}$ can lead to large changes in the overlaps. 
Second, thanks to the small energy splitting at the half-flux point, the non-adiabaticity required for the sudden-approximation description  \cite{sakurai} is readily achieved with pulse rising-edge times of $\lesssim 10\,\text{ns}$.
We fix the final flux value $\Phi_b$ away from half flux to a location where we can perform high-fidelity readout and where the charge matrix element is large enough between the lowest two levels to drive a fast $\pi$-pulse \cite{Zhu2013}. There is a range of flux values that satisfy these conditions below the $|1\rangle$-$|2\rangle$ state avoided crossing (see Fig.~ \ref{fig:FluxVsTime} near $\Phi = 0.85\ \Phi_0$). Here, we choose $\Phi_b = 0.812\ \Phi_0$. 
Measuring the occupation probability at the final flux value as a function of the starting flux values yields a range over which we can compare theory predictions with experimental data.

\section{Methods}
\label{sect:Methods}

\begin{figure*}[ht!]
    \centering
    \includegraphics[width=0.87\textwidth]{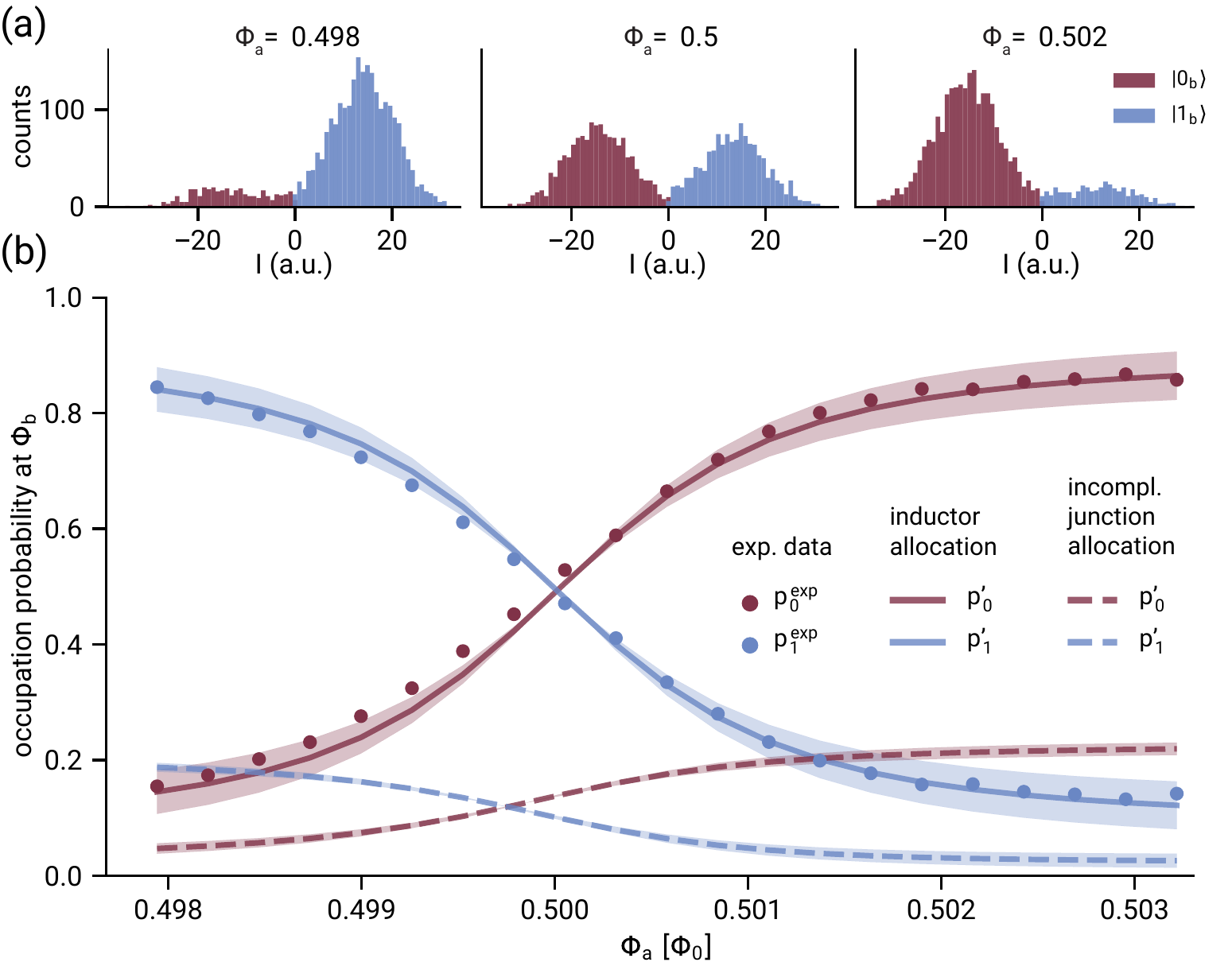}
    \caption{
        Occupation probabilities $p_m=|\langle m_b|0_a\rangle|^2$ after a sudden flux ramp from initial flux  $\Phi_a\in[0.498, 0.503]$ to final flux $\Phi_b = 0.812\ \Phi_0$, starting in  state $\keta{0}$. (a) Representative single-shot data of ground state (red) and first-excited state (blue) occupation probability when pulsed from $\Phi_a$ = 0.498, 0.5, 0.502 $\Phi_0$  to $\Phi_b = 0.812\ \Phi_0$.
        (b) Occupation probability as a function of initial flux point $\Phi_a$. Points represent measured data, solid lines signify inductor allocation, and dashed lines the incomplete junction allocation. The ground (first-excited) state is represented in red (blue). Experimental error bars are plotted on the data points, but are small enough to be enclosed within the size of the markers. Simulation curves are plotted accounting for measurement infidelity and state preparation errors as described in Sec.~\ref{sect:Methods}, with $\alpha = 0.05$ as the center and shaded region showing range $\alpha\in [0.0, 0.1]$. For the inductor allocation the occupation probability primarily remains in the qubit subspace, while for incomplete junction allocation most of the occupation probability would leak into higher-lying states (not shown). 
        }
    \label{fig:PopVsStartFlux}
\end{figure*}

In this work we use a 2D fluxonium circuit capacitively coupled to a co-planar waveguide (CPW) for both addressing and reading out the qubit \cite{Zhu2013}. The inductor is formed by a large array of 159 Josephson junctions. All junctions are made with Dolan bridges. To assist in reading out the qubit state we use a traveling-wave parametric amplifier \cite{White2015}. The external flux is controlled by both an on-chip flux line that runs near the loop of the fluxonium circuit and a global external magnet.  The global magnet is used to set the initial dc flux point $\Phi_{a}$, while the on-chip flux line is used to drive a step function flux pulse to the final point $\Phi_{b}$. The device is cooled and measured in a dilution refrigerator with a base temperature of $15\,\text{mK}$. We perform two-tone spectroscopy measurements \cite{Schuster2005}, and fit the data to determine the qubit parameters $E_{J}/h=6.49\,\text{GHz}$, $E_{C}/h=0.755\,\text{GHz}$, $E_{L}/h=0.445\,\text{GHz}$ and CPW cavity frequency of $7.39\,\text{GHz}$. At half flux, the frequency splitting between the ground and first-excited states is $20\,\text{MHz}$. Due to the low qubit frequency relative to the ambient temperature, both relaxation and excitation occur with nearly the same rate, with characteristic times $T_1 = 6.2 \pm 0.2$ $\mu$s. At the second flux point $\Phi = 0.812\ \Phi_0$, the qubit frequency is $5.018\,\text{GHz}$ and the coherence times are $T_1 = 47.3 \pm 2.3\,\mu$s and $T_{2R} = 72 \pm 7\,$ns. All readout pulses are 2\,$\mu$s long. Pulse generation and measurement are performed using the Quantum Instrumentation Control Kit (QICK) \cite{Stefanazzi2022}. 

For the sudden approximation experiment, we use the global magnet to set the initial dc flux point $\Phi_a$. Active state preparation is necessary due to the small energy splittings near half flux, leading to significant thermal population of the excited state. We perform single-shot readout \cite{Mallet2009, Gao2021,Riste2012} to post-select for the ground state at $\Phi_a$. The flux is quickly ramped to $\Phi_b$ by applying a step pulse to the on-chip flux line, see Fig.~\ref{fig:FluxVsTime} (b). The QICK generates the step pulse with a $1\,$ns rising edge. The experiment concludes with a final single-shot readout at $\Phi_{b}$ to determine the final-state occupation probability.

To quantify the measurement fidelity we perform a $\pi$-pulse calibration experiment at $\Phi_b = 0.812\ \Phi_0$. We use a $\pi$-pulse with a total length of 35 ns to drive into the excited state, identify the rotation angle to project the single-shot measurements onto a single axis with maximal contrast, and determine the threshold defining the ground and excited in the rotated axis. This calibration achieves a single-shot fidelity of 91\% which allows us to determine the measurement error rates for the ground and excited states. When measuring the ground (excited) state there is 5\% (4\%) population in the excited (ground) state. 

We incur state-preparation errors when operating near the half-flux point $\Phi_{a}$. Due to excitation processes, attempted initialization in the state $\keta{0}$ yields the mixed initial state
\begin{align}
\rho = (1-\alpha)\keta{0}\braa{0}+\alpha\keta{1}\braa{1},
\end{align}
where $0\leq\alpha\leq1$ denotes the preparation error. Based on the sudden approximation, the occupation probability in the state $\ketb{m}$ is
\begin{align}
\label{eq:prob}
p_{m}=(1-\alpha)|\bracketba{m}{0}|^2+\alpha|\bracketba{m}{1}|^2.
\end{align}
Based on the total measurement time of $2\,\mu$s, a qubit thermalization time of $6\,\mu$s, and our post-selection protocol we estimate $\alpha=0.05$. We present a range $\alpha \in [0, 0.1]$ for completeness. Additionally, we adjust the ground and excited-state occupation probabilities predicted by theory to account for the measurement infidelity from the calibration experiment:
\begin{align}
\left(\begin{matrix}
p_{0}' \\ p_{1}'
\end{matrix}
\right)=
\left(
\begin{matrix}
0.95 & 0.04 \\ 0.05 & 0.96
\end{matrix}
\right)
\left(\begin{matrix}
p_{0} \\ p_{1}
\end{matrix}\right).
\end{align}


\section{Results}

From the experimental data, we determine final occupation probabilities at $\Phi_b$ as a function of $\Phi_a$ and compare these results with the simulated estimates described in Sec.~\ref{sect:Theory}. Examples of single-shot measurements from starting external flux values of $\Phi_a$ = 0.498, 0.5, 0.502 $\Phi_0$ are shown in Fig.~\ref{fig:PopVsStartFlux}  (a). In these measurements the single shots are distributed into two groups which align with the ground and excited-state locations from the calibration experiment described in Sec.~\ref{sect:Methods}. 

We observe changes in occupation probability depending on the starting flux value $\Phi_a$.  
When below half-flux, there is a significantly greater occupation probability of the excited state compared with the ground state. By contrast when above half-flux, due to the change in the nature of the eigenstates the occupation probability of the ground state is greater than the first-excited state.
For the entire range of initial flux biases, the experimental data closely aligns with the predicted values obtained from the inductor allocation (or equivalently, the complete junction allocation retaining the flux-derivative term). In this allocation, the vast majority of the occupation probability ($>$98\%) is predicted to remain in the ground and first-excited states across all initial flux values $\Phi_a$. By comparison, the incomplete junction allocation predicts that the majority of the occupation probability escapes from the qubit subspace. We find that the incomplete junction allocation, omitting the time-derivative term, is inconsistent with the experimental data.

\section{Conclusion}

In this work we provide experimental verification of what the flux allocation should be in the presence of rapid time-varying flux in superconducting circuits. In particular, we perform experiments applying fast flux ramps to a fluxonium circuit, designed to test the theory predictions of You {\it et al.} \cite{You2019}. We initially bias the circuit at or near the half-flux point, then rapidly tune flux away to a common final value, and measure the qubit occupation probability. Our experimental results closely agree with the sudden-approximation predictions using the inductor allocation. Meanwhile, the incomplete junction allocation that erroneously omits the Hamiltonian term $\propto d\Phi/dt$ yields occupation probabilities which are inconsistent with the measured data. 
 
These results demonstrate that na\"ive omission of the flux derivative term leads to incorrect predictions. This constitutes experimental evidence supporting the theory for circuit quantization in the presence of time-dependent external flux proposed in Ref.~\cite{You2019}.  As flux modulation of multi-loop circuits becomes more prevalent, careful use of irrotational variables is essential to obtain correct predictions of device behavior.

\begin{acknowledgments}

We thank Jeronimo Martinez for helpful discussions of experimental techniques. D.~K.~W. acknowledges support from the Army Research Office (ARO) through a QuaCGR Fellowship. S.~S. is supported by the Department of Defense (DoD) through the National Defense Science \& Engineering Graduate Fellowship (NDSEG) Program. Research at Princeton University and Northwestern University was funded by the ARO under Grant No.\ W911NF-19-10016. Devices were fabricated in the Princeton University Quantum Device Nanofabrication Laboratory (QDNL) and in the Princeton Institute for the Science and Technology of Materials (PRISM) cleanroom.

\end{acknowledgments}

\bibliography{bibliography.bib}

\end{document}